\documentclass[a4paper]{article}

\usepackage{INTERSPEECH2022}
\usepackage{url}
\usepackage{graphicx}
\usepackage{amsmath, amsfonts}
\usepackage{bbm}
\usepackage{wasysym}
\usepackage{multirow}
\usepackage{xcolor}
\usepackage{tabu}

 \newfont{\msym}{msbm10}

\usepackage{siunitx}

\title{Self-supervised Speaker Diarization}
\name{Yehoshua Dissen$^1$, Felix Kreuk$^1$, and Joseph Keshet$^2$}
\address{
  $^1$Department of Computer Science, Bar-Ilan University, Ramat-Gan, Israel\\
  $^2$Electrical and Computer Engineering, Technion--Israel Institute of Technology, Haifa, Israel}
\email{shua.dissen@gmail.com, jkeshet@technion.ac.il}

\begin{document}

\maketitle

\begin{abstract}
Over the last few years, deep learning has grown in popularity for speaker verification, identification, and diarization. Inarguably, a significant part of this success is due to the demonstrated effectiveness of their speaker representations. These, however, are heavily dependent on large amounts of annotated data and can be sensitive to new domains. This study proposes an entirely unsupervised deep-learning model for speaker diarization. Specifically, the study focuses on generating high-quality neural speaker representations without any annotated data, as well as on estimating secondary hyperparameters of the model without annotations.

The speaker embeddings are represented by an encoder trained in a self-supervised fashion using pairs of adjacent segments assumed to be of the same speaker. The trained encoder model is then used to self-generate pseudo-labels to subsequently train a similarity score between different segments of the same call using probabilistic linear discriminant analysis (PLDA) and further to learn a clustering stopping threshold. We compared our model to state-of-the-art unsupervised as well as supervised baselines on the CallHome benchmarks. According to empirical results, our approach outperforms unsupervised methods when only two speakers are present in the call, and is only slightly worse than recent supervised models.  
\end{abstract}

% !TEX root =  main.tex

\section{Introduction}
\label{sec:intro}

Speaker diarization is the task of determining ``who spoke when'' in an audio recording, usually with an unknown number of speakers and variable speech duration. Traditionally, speaker diarization systems consist of four submodules. The first module extracts speech segments using a voice activity detector (VAD). The second module transforms the spoken segments into fixed-length vectors representing speakers' identity. This transformation is performed using a parametric probabilistic model, such as i-vectors based on GMM-UBM \cite{dehak2010front}, or using deep neural network (DNN) embeddings, such as x-vectors \cite{snyder2018x} or d-vectors \cite{wan2018generalized}. The third module finds a pairwise similarity measurement between segments using cosine distance, probabilistic linear discriminant analysis (PLDA) \cite{ioffe2006probabilistic}, or DNN models \cite{lin2019lstm}. Finally, the resulting similarity is clustered to determine the number of speakers and assign a speaker label to each segment. This is often done using a clustering model such as agglomerative hierarchical clustering (AHC) \cite{day1984efficient}, with a stopping threshold tuned on some held-out data \cite{fuchs2017spoken}.

One common theme amongst all the methods is the need for large amounts of labeled data, especially when considering DNN-based speaker embeddings. This would not be an issue if not for the domain sensitivity of the diarization systems. That is, training embeddings on one dataset might not translate into good representations on a different domain, and annotating data in the new domain can be prohibitively expensive. 

Our study aims to build the traditional diarization pipeline, but to replace each component with components that are trained completely unsupervised, thus removing the need for any annotated data. This opens new opportunities in applying speaker diarization especially for domains for which annotated data is not available. Hence, the work is focused on the main submodules of speaker diarization, namely, speaker embedding, segment similarity, and segment clustering, where the VAD is left out of the scope of this study.

In the self-supervised learning (SSL) setting, unlabeled inputs define an auxiliary task that can generate pseudo-labeled training data. Then, a model can be trained using supervised learning techniques, and typically with the contrastive loss function, which aims to learn a metric indicating that a pair of positive and negative examples are dissimilar. For example, the SSL model in \cite{kreuk2020self} is designed to distinguish between an adjacent frame, which is considered a positive example, and a randomly sampled future frame in the sequence, which is considered a negative example. 

In the supervised diarization setting, a positive and negative pair can correspond to the same speaker and different speaker segments, respectively. In the unsupervised setting, on the other hand, we have no supervision of who spoke when, hence we cannot easily generate positive and negative pairs. We cannot confidently assume that randomly sampled frames in an audio sequence represent a different speaker and hence considered as the negative frames as proposed in \cite{kreuk2020self}. Additionally, we cannot sample negative segments from different audio recordings as this can introduce other artifacts, and the negative examples might represent different acoustic environments rather than different speakers.

Due to these constraints, we take inspiration from recent developments in SSL for image representation \cite{zbontar2021barlow, bardes2021vicreg} that do not require negative examples. In those works, pairs of positive examples are generated by creating two distortions of the same image. The objective of the learning algorithm is to create a representation embedding invariant to distortions. In our setting, the inputs to the proposed models are raw speech segments considered to be spoken by the same speaker. Our setting aims to create a speaker-embedding that is invariant to the segment spoken by the speaker, and it is achieved by maximizing the agreement between embedding vectors of two different segments of the same speaker (rather than two distortions of the same speech segment.) 

The self-supervised speaker embeddings is then used to create a development set of pseudo-labels, which, in turn, are used to train a PLDA-based similarity matrix to the segments in a given audio stream. Additionally, the pseudo-labels are used to tune the speaker clustering stopping threshold. 

The contribution of this paper is a full unsupervised model for training a speaker diarization system, including all its sub-components: a speaker embedding model, a PLDA-based scoring matrix, and the detection thresholds. The results show that our unsupervised method is better than other unsupervised methods when two speakers are concerned, and comparable to the state-of-the-art fully-supervised methods from just a couple years ago. Our implementation is available at {\footnotesize \url{https://github.com/MLSpeech/ssl_diarization}}.

\smallskip 

\noindent {\bf Related work. }
There are few methods for unsupervised diarization that have been previously proposed. Castaldo \emph{et al.} \cite{castaldo2008stream} proposed an  expectation-maximization (EM) algorithm to the problem. First, a blind segmentation of the audio stream is performed. Then, a speaker model is estimated on the current segmentation to become the new global model or update the nearest speaker global model.
%is a GMM-UBM stream-based approach to speaker segmentation, where a sliding window of speech is segmented and, then each obtained speaker model is then paired with previous global models. 
Shum \emph{et al.} \cite{shum2013unsupervised} also proposed an iterative algorithm for speaker clustering based on principal component analysis and variational inference. Finally, the work  \cite{le2016speaker} proposes to use Bayesian Information Criterion (BIC) for creating training data and training a PLDA similarity metric on top of it. All those methods are based on i-vectors of GMM-UBM. Furthermore, they all need annotated data for tuning thresholds or biases. In contrast, our approach relies on recent advances in DNN-based speaker representation and does not use any annotated data for tuning thresholds or biases. 

% !TEX root =  main.tex

\section{Model}
\label{sec:model}

\begin{figure}[t]
  \centering
  \includegraphics[width=0.85\linewidth]{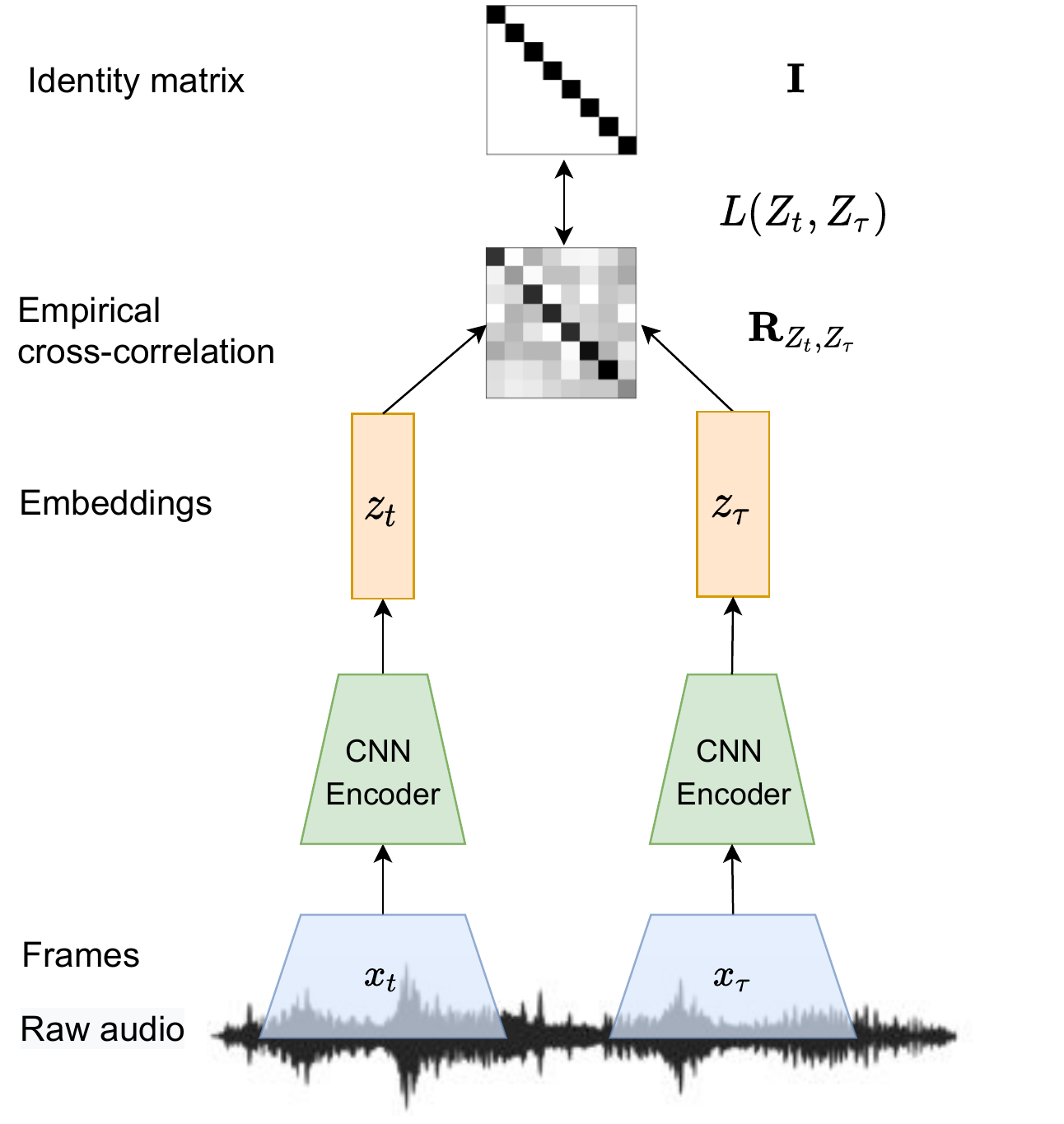}
  \caption{An illustration of our embeddings model and training scheme.}
  \label{fig:model}
\end{figure}

We denote a speech utterance of $T$ samples as $\mathbf{x}=\left(x_{1}, \ldots, x_{T}\right)$, where $\mathbf{x} \in \mathbb{R}^{T}$. We assume the existence of a label set $\mathcal{Y}=\left\{ \phi ,1,\ldots, K\right\}$, where $\phi$ is the symbol for silence or non-speech, and $K$ is the number of assumed speakers. In our setting, the value of $K$ is unknown at training time or during the inference, nor the labels associated with the inputs. We used the labels only for final evaluation purposes.

We denote by $\mathbf{x}_t=\mathbf{x}_{t-T'}^{t+T'}$ an audio segment of $2T'$ samples around the $t$-th sample, and we denote by $N$ the number of segments in the utterance. Our goal is to learn a function that predicts the speaker value  $y_t \in \mathcal{Y}$  for each segment, where it is assumed there is only one speaker within each segment. The proposed model is built from three components: an encoder $f_{\theta}$ with parameters $\theta$, a similarity scoring matrix $S\in\mathbb{R}^{N\times N}$ between the $N$ segments of the utterance, and a hierarchical clustering of the scoring matrix. 

The encoder $f_{\theta}$ is depicted in Figure~\ref{fig:model}. It gets as input a speech segment $\mathbf{x}_t$ and output an embedding vector $\mathbf{z}_t=f_{\theta}(\mathbf{x}_t)$. Ideally, the encoder should be trained using a contrastive loss function. However, since we assume we do not have access to segments of different speakers (negative examples), the encoder is trained using a non-contrastive self-supervised loss function. Our loss function gets two representations assumed to be of the same speaker as input. We denote these representations by $\mathbf{z}_t$ and $\mathbf{z}_{\tau}$ ($t\ne\tau$,) which corresponds to input segments $\mathbf{x}_t$ and $\mathbf{x}_{\tau}$, respectively. 

Denote the \emph{cross-correlation matrix} between the two embeddings as $\mathbf{R}_{\mathbf{Z}_t \mathbf{Z}_{\tau}}$:
\begin{equation}
\mathbf{R}_{\mathbf{Z}_t \mathbf{Z}_{\tau}} = \mathbb{E} \left[ \mathbf{z}_t^\top \mathbf{z}_\tau \right],
\end{equation}
where $\mathbf{Z}_t$ and $\mathbf{Z}_{\tau}$ denote the random variable associated with the vectors $\mathbf{z}_t$ and $\mathbf{z}_{\tau}$. Our self-supervised loss function is aimed at making the cross-correlation matrix $\mathbf{R}_{\mathbf{Z}_t \mathbf{Z}_{\tau}}$ as close as possible to the identity matrix, $\mathbf{I}$, under the Frobenius norm: 
\begin{equation}\label{eq:loss}
L_{BT}(\mathbf{z}_t, \mathbf{z}_{\tau}) =  \|\mathbf{R}_{\mathbf{Z}_t \mathbf{Z}_{\tau}} - \mathbf{I} \|^2_\mathcal{F} ~.
\end{equation} 
By using this loss, the diagonal elements of the matrix are steered towards 1 to make the embeddings invariant to different segments of the same speaker. In contrast, the off-diagonal elements are steered to 0, thus decorrelating the vector components of the embeddings. This loss function is known as the \emph{Barlow Twins} loss \cite{zbontar2021barlow}. Practically, the cross-correlation matrix is computed over the examples within the batch, and the parameters $\theta$ of the encoder function $f_\theta$ are found by minimizing this loss function on the unlabeled training set of utterances.

For subsequent processing, we generate a dataset with \emph{pseudo-speaker labels}. Our challenge is reliably labeling this dataset. First, we compute the embeddings of the utterances in our \emph{unlabeled development set} using the trained encoder. Then, we compute the Euclidean distances between these embeddings to generate the distance matrix $\mathbf{D}$ whose elements are:
\[
[\mathbf{D}]_{i,j} = \|\mathbf{z}_i - \mathbf{z}_{j}\|^2_2 ~.
\]
Next, we cluster the matrix $\mathbf{D}$ using agglomerative hierarchical clustering (AHC), resulting in a dendrogram, which is a diagram of the hierarchical relationship between segments. We decide on a cutoff point in the dendrogram that results in a number of clusters we assume to be higher than the speaker count of any file in our data, which consequently enables us to assume with high confidence cluster purity, i.e., the embeddings contained within each cluster -- all represent the same speaker. The higher the cutoff, the higher the confidence. Then, we take the embeddings from the largest-sized cluster from each utterance and label them as a pseudo-speaker. We repeat this process for each utterance in the development set, using a new pseudo-speaker label each time.

The second part of our system is a similarity scoring model $\mathbf{S}$, which outputs the similarity scores $\mathbf{S}(\mathbf{z})_{i,j}$ between $\mathbf{z}_i$ and $\mathbf{z}_j$ for all $i$ and $j$ segments of each utterance $\mathbf{z}$. The scoring model $\mathbf{S}$ is obtained by training a probabilistic linear discriminant analysis (PLDA) on the aforementioned development set in the same fashion as in the supervised setting, but with the pseudo generated labels \cite{prazak2011speaker}.

Finally, given an encoded utterance $\mathbf{z}$, we segment $\mathbf{z}$ to speakers by applying AHC on the similarity matrix $\mathbf{S}(\mathbf{z})$. 
For this, we need to select the threshold for the stopping criteria of the hierarchical clustering in an \emph{unsupervised} manner. We use our self-labeled development set to randomly sample segments (10 frames) from each pseudo-speaker (with repetitions). Next, we randomly concatenate these segments to create utterances with multiple speakers. We compute the similarity matrix for each generated utterance and use the $i$-th row of the similarity matrix $\mathbf{S}(\mathbf{z})$, denoted $\mathbf{S}(\mathbf{z})_i$, to represent the $i$-th segment of an input utterance. We set the similarity metric between segment $i$ and segment $j$ as $\|\mathbf{S}(\mathbf{z})_i-\mathbf{S}(\mathbf{z})_j\|^2$. Traditionally, the PLDA scores themselves are used for this clustering step. However, during our experiments, we found that using the row representations results in a more robust threshold than when merely using the pre-computed scores. Using these scores, we tune the clustering threshold on the development dataset and select the threshold that yields the lowest Diarization Error Rate (DER) on the pseudo labels. Empirically, we found that the resulting threshold also generalizes well to the test data.

% !TEX root =  main.tex

\section{Empirical Evaluation}
\label{sec:experimental_setup}

In this section we describe a set of experiments to demonstrate the efficiency of the proposed method. 

\subsection{Datasets} 
For training, we used Fisher English Training Speech Part 1 (LDC2004S13) and Part 2 (LDC2005S13), CallFriend, Switchboard-1 Release 2 \cite{godfrey1992switchboard}, RuSTeN (LDC2006S34), Gulf (LDC2006S43) and Iraq (LDC2006S45) Arabic Conversational Telephone Speech. Our development and test sets were different parts of CallHome. Supervised methods \cite{horiguchi2020end, zeghidour2021dive, fujita2019end, fujita2020end, horiguchi2021end, fujita2019end_} commonly reported Diarization error rate (DER) results on a subset of the CallHome with a development set consisting of 155 two-speaker calls, and a test set of 148 two-speaker calls prepared in \cite{fujita2019end_}. For a fair comparison we use the same subset in Table~\ref{tab:all_new} and Table~\ref{tab:ablations}. Unsupervised methods \cite{castaldo2008stream, shum2013unsupervised} reported DER on the full multi-speaker test set of CallHome (the NIST SRE 2000 set), which we also use in Table \ref{tab:shum}. Since the two-speaker development set is a subset of the full Callhome test, in this setting we used a different subset of the NIST SRE set, namely disks six and seven, where there are more than one speaker per call, as the development set.  

\subsection{Experimental setting}

We start with the speaker-discriminative embeddings training. The aforementioned training datasets were all converted into single-channel recordings. We applied VAD\footnote{We used Google's WebRTC with sensitivity 1 and frame size of 500 msec. It can be accessed at \url{https://github.com/wiseman/py-webrtcvad}.} to remove silences and non-speech. We randomly selected the inputs $\mathbf{x}_t$ and $\mathbf{x}_{\tau}$, where our segment size was 500 msec, and the segment pairs were 500 msec apart (using two adjacent segments resulted in an overfitting which did not generalize to good speaker separation). The encoder $f_\theta$ consists of 7 layers of 1-D convolutions (kernel sizes: 10,10,10,8,4,4,4; strides: 5,5,5,4,2,2,2; 128 channels per layer). Following this, there are 3 linear layers, each with 512 output units. The first 2 layers are followed by batch normalization and ReLU. We used the LARS optimizer \cite{you2017large}. We set a learning rate of 0.2 for the weights and 0.0048 for the biases and batch normalization parameters. We used a learning rate warm-up period of 10 epochs, after which we reduced the learning rate using a cosine decay schedule \cite{loshchilov2016sgdr}. The model was trained using a batch size of 2048. We trained the encoder $f_{\theta}$ until convergence (on our data, at around 30 epochs). This convergence also correlates well with an approximated DER we computed by clustering the produced embeddings. The resulting embeddings represent a 500 msec segment of speech.

To generate a self-supervised development set with pseudo labels, we took the development subset of CallHome and extracted embeddings using our encoder. Next, we clustered the embeddings using AHC. We cutoff the resulting dendograms arbitrarily at 10 clusters, selected the largest cluster and assumed cluster purity in the sense of it consisting of same speaker segments. 
% (here we used the number 10 arbitrarily).
We assumed each resulting cluster from different files corresponded to different speakers.
To test our assumptions regarding the clustering purity, we checked DER by labeling only the segments of the largest cluster as a single speaker in each file and marked the rest of the file as silence. We got only a 0.3\% speaker error, along with an extremely high percentage of missed speech which is to be expected. This validates our assumption that each of the selected clusters consists almost entirely of a single speaker, creating accurate pseudo-labels for a development dataset. 

The PLDA model was then trained on this pseudo labeled development set in the same fashion as in the supervised setting, using Kaldi's setup \cite{povey2011kaldi}. We also found a stopping threshold for the hierarchical clustering using this data by randomly generating 20,000 utterances. 

\subsection{Results}\label{sec:results}

\begingroup
\setlength{\tabcolsep}{3.3pt}
\begin{table}[t]
\caption{Speaker error rate on the multi-speaker test set of CallHome compared with unsupervised methods.}
\label{tab:shum}    
\centering
%\renewcommand{\arraystretch}{1.2}
%\small
    \begin{tabular}{lcccccccc}
    \hline
    \hline
    \#speakers & $2$ & $3$ & $4$ & $5$ & $6$ & $7$ & $avg$ \\ 
    \hline
    Castaldo \cite{castaldo2008stream} & $8.7$ & $15.7$ & $15.1$ & $20.2$ & $25.5$ & $29.8$ & $13.7$ \\
    Schum \cite{shum2013unsupervised} & $14.5$ & ${\bf 14.0}$ & ${\bf 13.0}$ & ${\bf 17.0}$ & ${\bf 19.5}$ & ${\bf 24.0}$ & $14.5$ \\
    %Ours & ${\bf 6.3}$ & $16.5$ & $22.6$ & $27.0$ & $41.1$ & $40.4$ & ${\bf 11.4}$
    Ours & ${\bf 6.1}$ & $14.9$ & $18.3$ & $23.0$ & $30.7$ & $30.5$ & ${\bf 13.4}$ \\
\hline
    \#calls & $303$ & $136$ & $43$ & $10$ & $6$ & $2$ & $500$\\
\hline
\hline
\end{tabular}
\end{table}
\endgroup

\begin{table}[t]
  \caption{Comparison of our unsupervised model with recent supervised methods. DER is reported on the two speaker test set of CallHome prepared in \cite{fujita2019end_}.
}
\label{tab:all_new}
\centering
\small
\setlength{\tabcolsep}{3pt}
\begin{tabular}{l S}
\hline
\hline
Model & {DER} \\
\hline
UIS-RNN V1  \cite{zhang2019fully} & 10.6 \\
UIS-RNN V2  \cite{zhang2019fully} & 9.6  \\
UIS-RNN V3  \cite{zhang2019fully} & 7.6  \\
x-vector + LSTM (oracle VAD) \cite{lin2019lstm} & 6.6 \\
DIVE  \cite{zeghidour2021dive} &  5.9 \\
\hline
Ours (unsupervised, oracle VAD)   &  { 6.6} \\
Ours (unsupervised) &   { 9.1} \\
\hline
\hline
\end{tabular}
\end{table}

We turn to evaluate our method against state-of-the-art unsupervised and supervised models. Then, we demonstrate the effect of the amount of (unsupervised) data on the system's performance. Finally, we conclude with a detailed ablation study to gauge the impact of each component of our system. All measurements were obtained using the NIST DER tool with a collar of 250 msec and ignore overlapping sections.

We start by presenting our work compared with other unsupervised works mentioned in Section \ref{sec:intro}.
In these papers, the full multi-speaker test set of CallHome was used. This amounts to 500 recordings, each 2-5 minutes in length, containing between 2 and 7 participants.

Table \ref{tab:shum} reports the confusion error per speaker count and the average speaker error rate using oracle VAD of our work compared with two unsupervised methods: Castaldo \emph{et al.} \cite{castaldo2008stream} and Shum \emph{et al.} \cite{shum2013unsupervised}.
Since an oracle VAD is employed in these measurements, we
exclude false-alarms and miss-detection from our evaluations, and report speaker confusion only.
Our system is better for the two speakers setting and our overall (weighted) results are better than both methods.

Next, we would like to compare the performance of our unsupervised system to the performance of state-of-the-art supervised systems from recent years. Note that supervised systems are often evaluated on the two-speakers subset of the dataset that used in the evaluation of unsupervised methods. Table \ref{tab:all_new} reports DER on the two speakers test set of CallHome compared to recent state-of-the-art \emph{supervised} methods. UIS-RNN \cite{zhang2019fully} is a hybrid system training an RNN on top of pre-trained speaker embeddings, with the three version: V1 was trained on 36M utterances from 18K US English speakers (all mobile phone data); V2 was trained on additional non-public containing 34M utterances from 138K speakers, LibriSpeech, VoxCeleb, and VoxCeleb2, and V3 was based on variable-length processing windows, drawn uniform at random from [240 msec, 1600 msec]. DIVE \cite{zeghidour2021dive} combines three modules which are trained jointly: projection of the waveform to an embedding space, iterative selection of long-term speaker representations, and per-speaker per-timestep VAD and is the current state of the art reaching 5.9\% without overlap. Similar to UIS-RNN, \cite{lin2019lstm} suggested training an LSTM to model the similarity between pretrained speaker embeddings and then to perform the diarization. Unlike the other models in Table \ref{tab:all_new}, it uses oracle speech activity labels (removing silences). 

The last rows are the result of our model, with learned embeddings, PLDA, and clustering tuned on pseudo labels, with and without overlap. The error breakdown without overlap is as follows: 1.5\% missed-speech, 2.2\% false-alarm, and 5.2\% speaker error. For a fair comparison, we added the performance of our system against using the oracle VAD labels. In this setting, we achieve a 6.58\% DER comparable to their 6.63\%. As seen, our unsupervised system is only slightly worse than state-of-the-art supervised systems from recent years. 

\begin{figure}[t]
  \centering
  \includegraphics[width=0.95\linewidth]{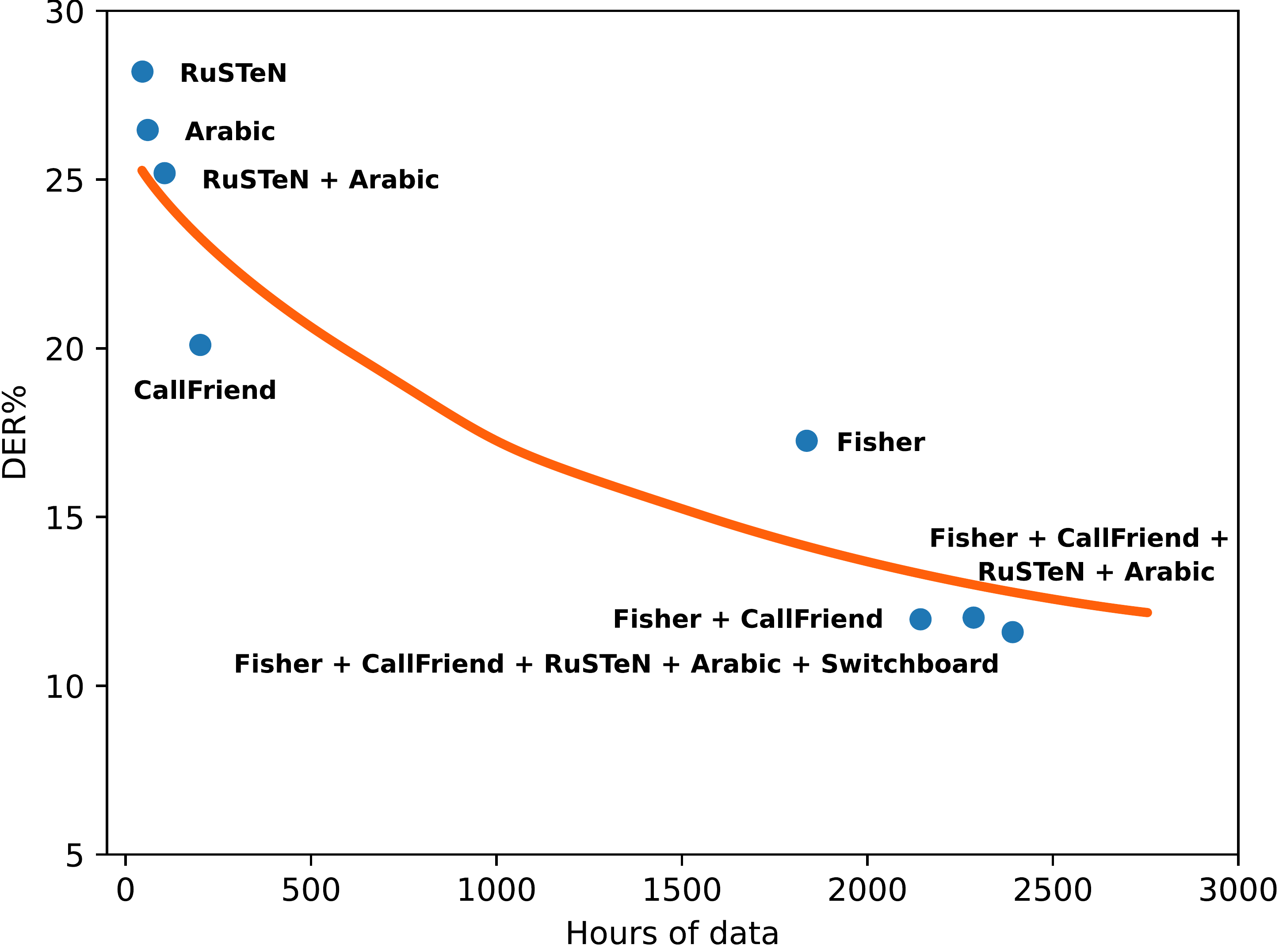}
  \caption{Drop in DER as a function of the amount of hours of training data used for training.}
  \label{fig:data}
\end{figure}

In Figure \ref{fig:data} we show that the more unlabeled data that is used, the better the embedding model. Here we present different combinations of datasets we used and the total hours of training data used in each mode. The total hours are calculated by summing all the VAD segments longer than 3 seconds, as this is what was used for training and the DER is the calculated by clusering the embeddings using the number of speakers as a stopping criteria. As seen the DER ranges from 28\% for 50 hours of data to 11.5\% for 2400 hours. This bodes well for the system as most diarzation users will have access to large amounts of in domain unlabeled data.

\iffalse
\begin{table}
  \caption{An ablation study of the different system components. Results are reported in terms of DER [\%] on the two-speaker test set of CallHome prepared in \cite{fujita2019end_}.}
  \label{tab:ablations}
    \centering
    \small
    \setlength{\tabcolsep}{3pt}
  \begin{tabular}{l l l}
    \hline
    \hline
   Model & {Overlap} & {No overlap} \\
    \hline
    Embeddings + clustering + \#spk            &  16.94  & 12.03 \\
    Embeddings + PLDA + clustering          &  14.47  & 9.15 \\
    Embeddings + PLDA + \#spk &  14.21      &  8.89 \\
    Embeddings + PLDA + clust. + (oracle VAD) &    12.31   &  6.59 \\
    VicReg embeddings + clustering + \#spk   &  17.9     &  13.1 \\
    \hline
    \hline
  \end{tabular}
\end{table}
\fi

\begin{table}
  \caption{An ablation study of the different system components. Results are reported on the two-speaker test set of CallHome.}
  \label{tab:ablations}
    \centering
  \begin{tabular}{l S}
    \hline
    \hline
   Model & {DER} \\
    \hline
    Embeddings + clustering + \#spk & 11.5 \\
    Embeddings + PLDA + clustering & 9.1 \\
    Embeddings + PLDA + \#spk & 8.7 \\
    Embeddings + PLDA + clust. + (oracle VAD) & 6.6 \\
    VicReg embeddings + clustering + \#spk & 13.1 \\
    \hline
    \hline
  \end{tabular}
  \vspace{-.3cm}
\end{table}

Last, we evaluated our system by methodically omitting the different components. The results in terms of DER [\%] on the test set of CallHome are reported in Table \ref{tab:ablations}. 

In the table, \emph{Embeddings + clustering + \#spk} denotes the results when using only the learned speaker-embeddings without incorporating the learned PLDA similarity matrix, but where the speaker count is known and is used as a stopping criterion for the clustering procedure. This is essentially a measure of the goodness of the learned speaker embeddings. \emph{Embeddings + PLDA + clustering} refers to our final choice of architecture, as described in Section~\ref{sec:model}.

\emph{Embeddings + PLDA + \#spk} is our final model when the speaker count in each file is known and used as the clustering stopping criteria. As expected, knowing the speaker count in each file yields slightly better results. These results suggest that incorporating our proposed self-supervised PLDA similarity matrix improved results by 24\% in relative terms over pure embeddings. \emph{Embeddings + PLDA + oracle VAD} denotes the results when using the oracle VAD. This shows that improving the VAD can significantly lower the DER. 

For completeness, we also trained the speaker-embedding system using the \emph{VicReg} objective proposed in \cite{bardes2021vicreg}. This objective similar to \emph{Barlow Twins}, exchanges the variance term with a hinge loss function and adds a term minimizing the MSE between embedding vectors.  We denote this setting by \emph{VecReg embeddings + clustering + \#spk}. However results suggest that this system was inferior by approximately 1.5\% absolute DER.

% !TEX root =  main.tex

\section{Discussion and future work}
\label{sec:conclusions}

This paper introduces a new approach for completely unsupervised speaker diarization. We showed that this method with a relatively small model, performs close to state-of-the-art fully supervised methods on the CallHome test set, further closing the gap between supervised and unsupervised methods. 

For future work, we will explore ways of dealing with overlapping speech, which is not referenced in this work. Additionally, we will examine ways to incorporate negative examples in a way that can further improve results. Finally, we are interested in studying how leveraging massive amounts of unlabeled data affects this model.

\bibliographystyle{IEEEtran}
\bibliography{refs}

\end{document}